\documentclass[pra,letterpaper,aps,10pt,superscriptaddress,twocolumn,floatfix,showpacs]{revtex4-1}
\usepackage{amsmath,graphicx,amssymb,braket,xcolor,subfigure,upgreek}

\begin{document}

\title{Protected subspace Ramsey spectroscopy}
\author{L. Ostermann, D. Plankensteiner, H. Ritsch and C. Genes}
\affiliation{Institut f\"ur Theoretische Physik, Universit\"at
Innsbruck, Technikerstrasse 25, A-6020 Innsbruck, Austria}
\date{\today}

\begin{abstract}
We study a modified Ramsey spectroscopy technique employing slowly decaying states for quantum metrology applications using dense ensembles. While closely positioned atoms exhibit superradiant collective decay and dipole-dipole induced frequency shifts, recent results [Ostermann, Ritsch and Genes, Phys. Rev. Lett. \textbf{111}, 123601 (2013)] suggest the possibility to suppress such detrimental effects and achieve an even better scaling of the frequency sensitivity with interrogation time than for noninteracting particles. Here we present an in-depth analysis of this 'protected subspace Ramsey technique' using improved analytical modeling and numerical simulations including larger 3D samples. Surprisingly we find that using sub-radiant states of $N$ particles to encode the atomic coherence yields a scaling of the optimal sensitivity better than $1/\sqrt{N}$.  Applied to ultracold atoms in 3D optical lattices we predict a precision beyond the single atom linewidth.
\end{abstract}


\maketitle

\section{Introduction}

Recent experimental setups have demonstrated Raman and Ramsey spectroscopy on narrow atomic clock transitions using cold atoms trapped in 1D magic wavelength optical lattices with unprecedented precision below one Hertz~\cite{akamatsu2014spectroscopy,bloom2014optical}. While on the one hand in this extreme limit even weak atom-atom interactions cause perturbations, on the other hand such setups provide a unique testing ground for measuring such tiny corrections~\cite{rey2014probing}. From the point of view of an atomic clock or a superradiant laser~\cite{maier2014superradiant} interactions constitute a perturbation. In particular at higher particle densities dipole-dipole interaction and collective decay tend to introduce  shifts and dephasing~\cite{lehmberg1970radiation,ficek2002entangled,ostermann2012cascaded,zoubi2010metastability}, which limit the useful interrogation time. As these are essentially bipartite interactions, they cannot be corrected simply by rephasing techniques. While for $^{87}Sr$ with its mHz-linewidth, decay is no major limitation at the moment, alternative approaches with e.g. calcium atoms have already reached this limit~\cite{Oates2014atomic}.

Over the past couple of years a considerable number of theoretical proposals to deal with metrology bounds have been put forward (see Refs.~\cite{xu2014conditional, zhang2014precision, dur2014improved, zhang2014quantum, wang2014heisenberg, dorner2013noise}). In our recent theoretical proposal~\cite{ostermann2013protected} we suggested that by a proper modification of the standard Ramsey interferometry technique (SRT) on interacting two-level ensembles, the detrimental effect of collective decay can be minimized and surprisingly to some extent even reversed. The technique takes advantage of the atomic interactions to suppress decay by transferring the atomic excitation to subradiant collective states. We dub this method protective Ramsey technique (PRT). It might be less surprising in hindsight, but still is puzzling, that an optically highly excited collective state of atomic dipoles can be prevented from decay via destructive interferences of the field emitted by the individual dipoles. Interestingly, one finds an unexpected fast growth of the lifetime of the excited states with the particle number. Employing the proposed techniques these long-lived states can then be used for an enhanced Ramsey spectroscopy allowing for a significantly higher precision than even for independently decaying atoms paving the way for implementations of this technique with 3D lattices.

The method requires an additional individually controlled single particle spin rotation, which is added after the first and reversed before the final Ramsey pulse. In consequence, the total ensemble spin is shifted towards zero by spreading the individual spins by predefined amounts almost homogeneously around the equatorial plane of the Bloch sphere. Thus the ensemble becomes classically nonradiative during free evolution. While this should obviously work for tightly packed ensembles confined within a cubic wavelength, we demonstrate that it works almost as well in 3D regular lattices. In this case it is not a priori clear which would be the most long lived configuration, but the minimum decay rate can be inferred from the eigenvalues of the collective decay Liouvillian operator. It is of course an extra technical challenge to implement the required optimal transformation as it  in general requires individual spin addressing. In practice, however, in many cases, a proper use of phases introduced by a designed lattice and excitation geometry turns out to be sufficient to get very close to such an optimal state with a single laser applied at an optimal angle.

It is generally thought that, in order to beat the $1/\sqrt{N}$ scaling of the sensitivity of SRT applied on $N$ noninteracting particles, the state preparation stage should involve the generation of nonclassical multipartite entangled states (such as spin squeezed states)~\cite{wineland1992spin,meiser2008spin, oblak2005quantum,louchet2010entanglement,borregaard2013near,leroux2010focus,leroux2010orientation}. Here we present numerical evidence that suggests that one can overcome this scaling by employing classical operations at the initial and final stages of the sequence only.

In Sec. II we describe our model and discuss the formalism, while Sec. III gives an overview of the results of SRT applied to non decaying or independently decaying atoms. We introduce PRT in Sec. IV and elaborate on our choice of rotations. We also detail the method applied to simple interacting systems comprised of two atoms and three atoms in a triangular geometry. The main body of numerical results and analytical considerations for larger systems is presented in Sec. V, where
chains of many atoms are considered, scaling laws are investigated and results for the fundamental cubic unit cell are presented. We conclude in Sec. VI.

\section{Model}
 Our model assumes $N$ identical two-level atoms with levels $\left \vert g\right\rangle$ and $\left\vert e \right\rangle$ separated by an energy of $ \hbar \omega_0 $ (transition wavelength $\lambda_0$) in a geometry defined by the position vectors $ \left \lbrace \mathbf{r}_i \right \rbrace$ for $i=1,...N$. For each $i$, operations on the corresponding two-dimensional Hilbert space are written in terms of the Pauli matrices $\sigma_i^{x,y,z}$ and corresponding ladder operators $ \sigma_i^\pm$ connected via
\begin{subequations}
\begin{align}
\sigma_i^x &= \sigma_i^+ + \sigma_i^- \\
\sigma_i^y &=-i(\sigma_i^+ -\sigma_i^-) \\
\sigma_i^z &=  \sigma_i^+\sigma_i^- - \sigma_i^- \sigma_i^+.
\end{align}
\end{subequations}

Rotations about an axis $\mu$ are defined as
\begin{equation}
\mathcal{R}^{(j)}_\mu [\varphi] = \exp \left( i \varphi \, \sigma^\mu_j / 2 \right),
\end{equation}
where $\mu \in \{ x, y, z \}$. The coupling of the system to the common bath represented by the surrounding electromagnetic vacuum results in i) irreversible dynamics characterized by independent decay channels with rates $\Gamma_{ii}\equiv\Gamma$ as well as cooperative decay channels with rates $\Gamma _{ij}$ (for atom pair $\{i,j\}$) and ii) dipole-dipole interactions through the exchange of virtual photons characterized by the frequency shifts $\Omega _{ij}$. Assuming identical dipole moments for all atoms, we can write this explicitly~\cite{ostermann2012cascaded} as
\begin{subequations}
\begin{align}
\Omega_{ij} &= \frac{3 \Gamma}{4} \, G( k_0 r_{ij}) \\
\Gamma_{ij} &= \frac{3 \Gamma}{2} \, F( k_0 r_{ij})
\end{align}
\end{subequations}
for two atoms separated by a distance of $r_{ij}$. With the notations $\xi = k_0 r_{ij}$ (with the wavenumber $k_0=2\pi/\lambda_0$) for the normalized separation and $\alpha = \cos \theta = \left(\mathbf{r}_{ij} \cdot \mathbf{\mu} \right) / |\mathbf{r}_{ij} || \mathbf{\mu} |$, one can put down the two functions
\begin{subequations}
\begin{align}
F \left( \xi \right) =& \left( 1- \alpha^2 \right) \frac{\sin \xi }{\xi} \\
&+ \left( 1- 3 \alpha^2 \right) \left( \frac{\cos \xi}{\xi^2 }- \frac{\sin \xi}{\xi^3} \right),\\
G \left( \xi \right) =& - \left( 1- \alpha^2 \right) \frac{\cos \xi}{\xi} \\
&+ \left( 1- 3 \alpha^2 \right) \left( \frac{\sin \xi}{\xi^2}+\frac{\cos \xi}{\xi^3} \right).
\end{align}
\end{subequations}

We follow the evolution of the system both analytically and numerically in the framework of the master equation
\begin{equation} \label{master} \frac{\partial \rho}
{\partial t}=i[\rho ,H]+\mathcal{L}[\rho ].
\end{equation}
The unitary dynamics of the system is described by the Hamiltonian
\begin{equation}
H = \frac{\omega}{2} \sum_{i} \sigma _i^z + \sum_{i \neq j}\Omega_{ij} \, \sigma _i^+ \sigma_j^-
\end{equation}
with $\omega=\omega_0-\omega_l$, where $\omega_l$ is a laser reference frequency. The dissipative dynamics can be written in (a nondiagonal) Lindblad form
\begin{equation} \label{L_def}
\mathcal{L}[\rho ]= \frac{1}{2} \sum_{i,j} \Gamma _{ij} \left[ 2 \sigma_i^- \rho \, \sigma_j^+ -\sigma_i^+ \sigma_j^- \rho -\rho \, \sigma_i^+ \sigma_j^- \right].
\end{equation}

\section{Standard Ramsey Interferometry}

\begin{figure}[t]
 \includegraphics[width=0.99\columnwidth]{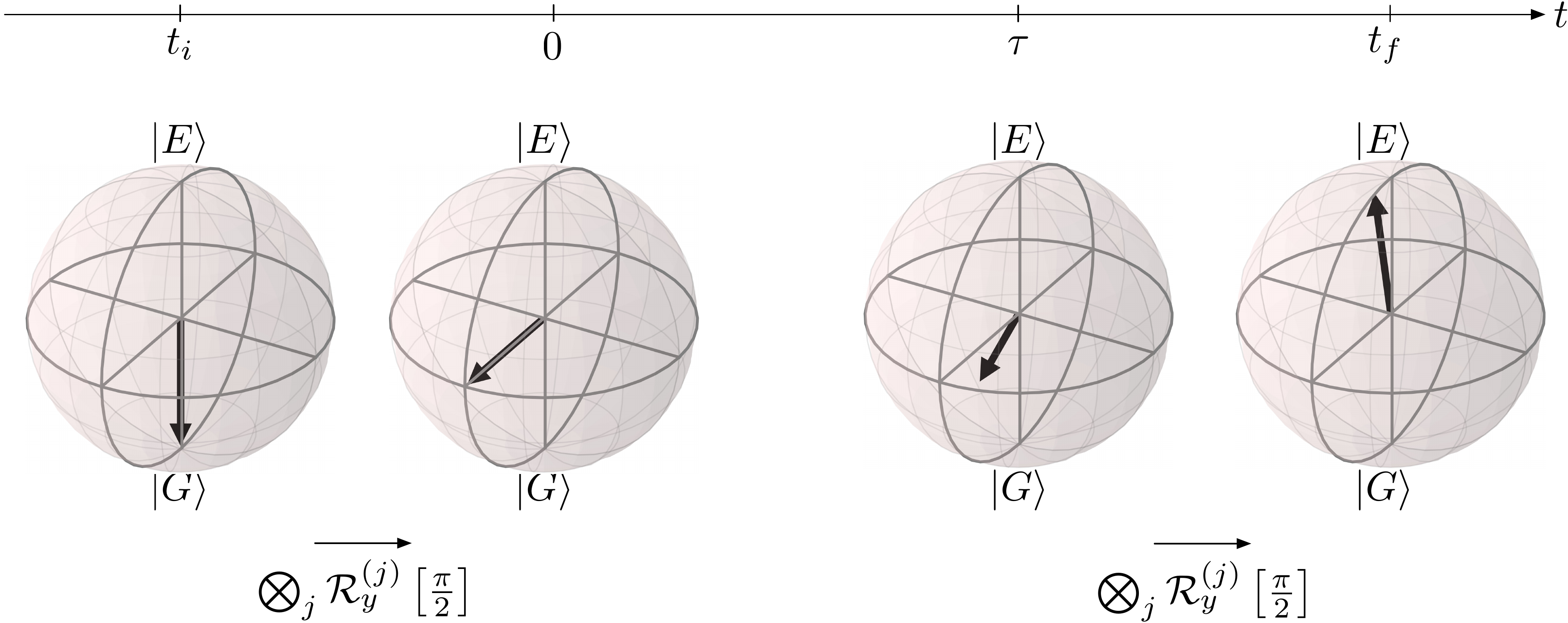}
 \caption{\emph{Standard Ramsey metrology}. The ensemble of $N$ spins starts with all spins down in a collective coherent pure spin state on the surface of the collective Bloch sphere (radius $N/2$). The first $\pi/2$ pulse aligns the average collective dipole along the $x$ axis and free evolution is allowed. After the interrogation time $\tau$, another $\pi/2$ pulse follows which attempts to align the state with the excited state and fails by an angle that depends on the accumulated phase during free evolution as well as on the total decay of the collective state. The detected signal to be analyzed is a measure of population inversion.}
 \label{fig1}
\end{figure}

Let us review some fundamental aspects of a typical procedure in spectroscopic experiments, i.e. the Ramsey method of separated oscillatory fields~\cite{Ramsey1990molecula}. As illustrated in Fig.~\ref{fig1}, the method consists of preparing an ensemble of
spins in the ground state at time $t_i$ such that their collective population $S^z= \sum_i \sigma_i^z /2$ starts at a value of $\langle S^z \rangle =  -N/2$. A preparatory Ramsey pulse, applied between $t_i$ and $t=0$, rotates the state around the $y$-direction to achieve an alignment of the collective dipole with the $x$-axis. This is realized by applying a laser that is quasi resonant with the atomic transition with a Rabi frequency $\chi$ for the time $t_i-t_0$ such that the pulse area $\int_{t_0}^{t_i} \, \chi(t^\prime) \, \mathrm{d}t^\prime \approx \pi/2$. As a simplification we assume that $\Omega_{ij},\Gamma_{ij} \ll \chi$ such that no population redistribution among the atoms can occur during the pulse. Typically, for level shifts and decay rates on the order of MHz, a Rabi frequency in the GHz regime or more would ensure that this approximation is valid for laser pulses with a duration in the realm of ns. In the next step, the ensemble is allowed to evolve freely for what we refer to as 'interrogation time' $\tau$. Note that, depending on the geometry of the excitation scheme (whether the laser comes from the side or propagates through the ensemble) the signal will show oscillations in time either at laser-atom detuning $\omega$ or at the natural frequency $\omega_0$. The next step is the same as the first one, where a second $\pi/2$ pulse rotates the collective state around the $y$-axis. At the end, the signal to be extracted is the population inversion as a function of the scanned laser detuning. Analysis of this signal gives the sensitivity as a
figure of merit in metrology
\begin{equation}\label{sensdef}
\delta \omega =\min \left[\frac{\Delta S^z (\omega,\tau)}{\left|
\partial_{\omega}\langle S^z\rangle (\omega,\tau) \right| }\right],
\end{equation}
where the minimization is performed with respect to $\omega$ and $S^z = \sum_i \sigma^z_i$ is the detected signal, while $\left( \Delta S^z \right)^2= \left \langle \left( S^z \right)^2 \right \rangle - \langle S^z \rangle^2$ refers to its rms deviation.

To start with, we assume independent systems ($\Gamma _{ij} = 0$ and $\Omega _{ij} = 0$ for $i\neq j$). The operations to be applied on the density matrix $\rho$ at the times of the Ramsey pulses are
\begin{equation}
\mathcal{R}_1 = \mathcal{R}_2 = \bigotimes_j \mathcal{R}^{(j)}_y
[\pi/2].
\end{equation}

It is easy to find the optimal sensitivity as a minimization over $\omega$ as
\begin{equation} \label{sens-indep}
\left[ \delta\omega \right]_{\text{indep}} = \min \left[{\frac{\sqrt{e^{\Gamma \tau} - \cos^2(\omega \tau)}}{\sqrt{N} \left| \tau \cdot \sin(\omega \tau) \right| }}\right]= \frac{e^{\Gamma \tau/2}}{\tau \sqrt{N}}.
\end{equation}

Notice that, for nondecaying atomic excitations, the method allows for a perfect accuracy,. However, in the presence of decay, an optimal interrogation time $\tau_{opt}=2/\Gamma$ suggests itself, where the corresponding optimal sensitivity is given by
\begin{equation} \label{indep-min-sens}
\left[ \delta \omega \right]_{\text{indep}}^{\text{opt}} =
\frac{\Gamma \cdot e}{2\sqrt{N}}.
 \end{equation}
 
Thus, it becomes obvious that, given the atomic species (which determines $\Gamma$) one can improve the accuracy by an increase of the sample size only. Yet, due to the finite available volume, this would imply an increase of density which causes the assumption that the atoms are independent to break down. In the next section we analyze this high density limit where we observe that the collective behavior can be exploited to reduce the effective $\Gamma$ appearing in  Eq. \ref{indep-min-sens} instead.

\section{Protective Ramsey Technique}

\begin{figure}[t]
 \includegraphics[width=0.98\columnwidth]{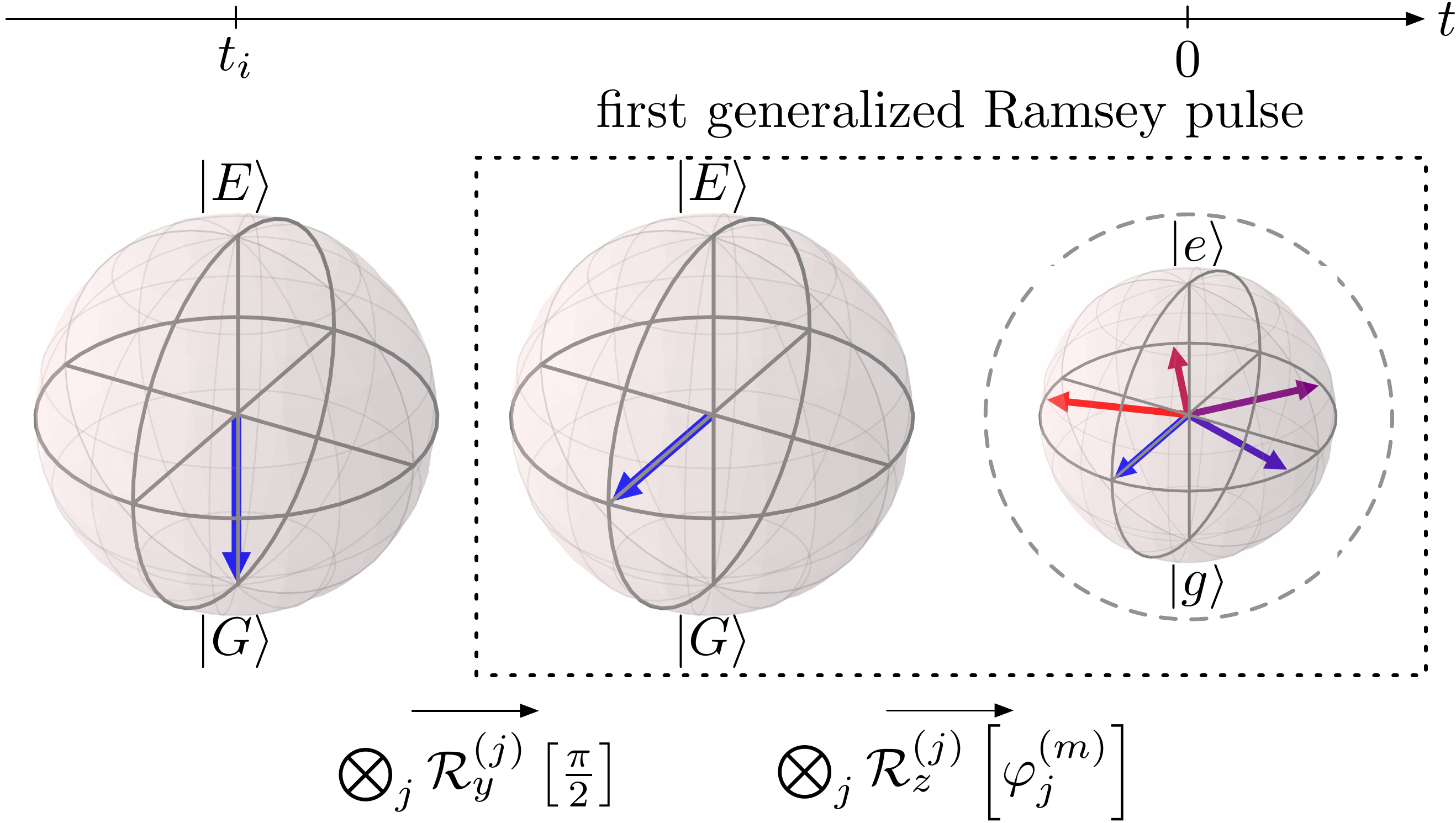}
 \caption{\emph{Phase-spread operation}. Redesign of the first Ramsey pulse to include, in addition to the initial $\pi/2$ pulse, rotations of the individual spins with different angles such that the resulting overall dipole moment vanishes. Notice that the last Bloch sphere has a  radius $1/2$, corresponding to single spins.}
 \label{fig2}
\end{figure}

To counteract the effect of the collective coupling to the vacuum modes, it has been proposed~\cite{ostermann2013protected} to make use of a generalized Ramsey sequence (as illustrated in Fig.~\ref{fig2}). In contrast to SRT, the generalized PRT contains extra rotations in conjunction with the Ramsey pulses that are intended to drive the spin system into states that are protected from the environmental decoherence. In a first step, one applies
 \begin{equation}
  \mathcal{R}_1^{(m)} = \bigotimes_j \mathcal{R}^{(j)}_z \left[ \varphi_{j}^{(m)} \right] \cdot \mathcal{R}^{(j)}_y \left[ \frac{\pi}{2} \right],
\end{equation}
where the state of a particular atom $j$ is rotated around the $z$-direction with the angle
\begin{equation} \label{phi}
\varphi_{j}^{(m)}=2\pi m \,  \frac{j-1}{N}.
\end{equation}

The idea behind this choice of angles is to drive the system into a state which exhibits a vanishing dipole moment. This can be achieved
by rotating the spins in the $xy$-plane in $\lfloor N/2 \rfloor$ distinct ways indexed by $m=1,...\lfloor N/2 \rfloor$ (where $\lfloor N/2 \rfloor$ is the first integer before $N/2$). The
protection of the state is targeted at the period of its free evolution and in the final step, before the second Ramsey pulse, the state has to be brought back to the surface of the Bloch  sphere to ensure a large contrast in the signal. This is accounted for by a reversal of the phase spread operation, i.e.
\begin{equation}
   \mathcal{R}_2^{(m)} = \bigotimes_j \mathcal{R}^{(j)}_y \left[ \frac{\pi}{2} \right] \cdot \mathcal{R}^{(j)}_z \left[- \varphi_{j}^{(m)} \right].
\end{equation}

As  stated previously, at time $t=0$, for any set of $\varphi_{j}^{(m)}$, the system is in a state of zero average collective spin. At an intuitive level this choice comes from the observation that, for small atom-atom separations, collective states of higher symmetry are shorter lived (culminating at zero separation with the maximally symmetric superradiant Dicke state~\cite{dicke1954coherence} of rate $N \Gamma$). Let us now try to sketch how asymmetric states can be built by imposing orthogonality of a phase-spread state
\begin{equation}
\left| {\psi_\varphi} \right \rangle = \bigotimes_{j = 1}^{N} \frac{1}{\sqrt{2}} \, \left[ \ket{g} + \left( e^{i \varphi} \right)^{(j-1)} \ket{e} \right],
\end{equation}
to the multitude of symmetric states of the system. It is straight-forward to see that
\begin{equation}
\left \langle W \vert \psi_\varphi \right \rangle = \sum_{i=1}^N \left( e^{i \varphi} \right)^{(j-1)},
\end{equation}
where $\left \vert W \right \rangle = \left( \left \vert e g g \dots
\right \rangle + \left \vert g e g \dots \right \rangle + \dots +
\left \vert g \dots g e \right \rangle \right)/\sqrt{N}$, the
so-called $W$-state, which is the fully symmetric state of a single excitation distributed equally among $N$ atoms. Imposing orthogonality, i.e. $\left \langle W \vert \psi_\varphi \right
\rangle = 0$ we find $\varphi = 2 \pi /N$. Geometrically, this corresponds to a division of the unit circle into $N$ pieces of angle $2 \pi /N$, which when added up yields a trivial vector sum of zero. Generalizing this concept to higher energy states, where $\left \vert w^{(n)} \right \rangle$ is the symmetric state of $n$ excitations, gives us
\begin{equation}
\left \langle w^{(n)} \vert \psi_\varphi \right \rangle = \sum_{j = 1}^M p(j, n) \, \left( e^{i \varphi} \right)^{(j-1)} = 0
\label{wn}
\end{equation}
with $p(j, n)$ being the integral partition of the number $j$ comprised of $n$ summands and $M = n \left(N-(n+1)/2 \right)+1$. Unfortunately, $p(j, n)$ is a fractal function and thus, Eq.~\eqref{wn} cannot be solved for a general number of atoms and excitations, yet any concrete number gives the same result as above, i.e. $\varphi = 2 \pi / N$. Hence, we see that for any symmetrically coupled system of $N$ atoms the choice $\varphi = 2 \pi /N$ results in a zero-occupation of the symmetric states.

We will now look at systems of small atom numbers where the protected states $\ket{p_N^{(m)}}=\mathcal{R}_1^{(m)}\ket{G}$, with $\ket{G}$ being the ground state, can be readily expressed in both the collective and uncoupled bases. For two atoms the 'protected' state is unique ($m=1$) and is simply the asymmetric state
\begin{equation}
\ket{p_2^{(1)}} = \frac{1}{\sqrt{2}} (\ket{ge}-\ket{eg})=\ket{A}.
\end{equation}
Observe that the transformation that diagonalizes the Hamiltonian automatically renders the Liouvillian in diagonal form. Denoting the mutual decay rate by $\gamma_{12} = \gamma$, two decay channels with $\gamma_A=\Gamma-\gamma$ and $\gamma_S=\Gamma+\gamma$ are obtained. For closely spaced atoms, $\gamma$ can reach values close to $\Gamma$ such that $\gamma_A\ll\Gamma$ and the state $\ket{A}$ can be protected from decoherence very well. Since analytical and numerical results for the two atom case are presented in Ref.~\cite{ostermann2013protected}, we will only stress one conclusion that emerges from this analysis, i.e., even for moderate distances the time for which the optimal sensitivity is obtained roughly scales as $2/\gamma_A$. This indicates that the evolution of the system is mainly within the protected subspace, a claim that will be investigated further in the next section.

\begin{figure}[t]
 \includegraphics[width=0.68\columnwidth]{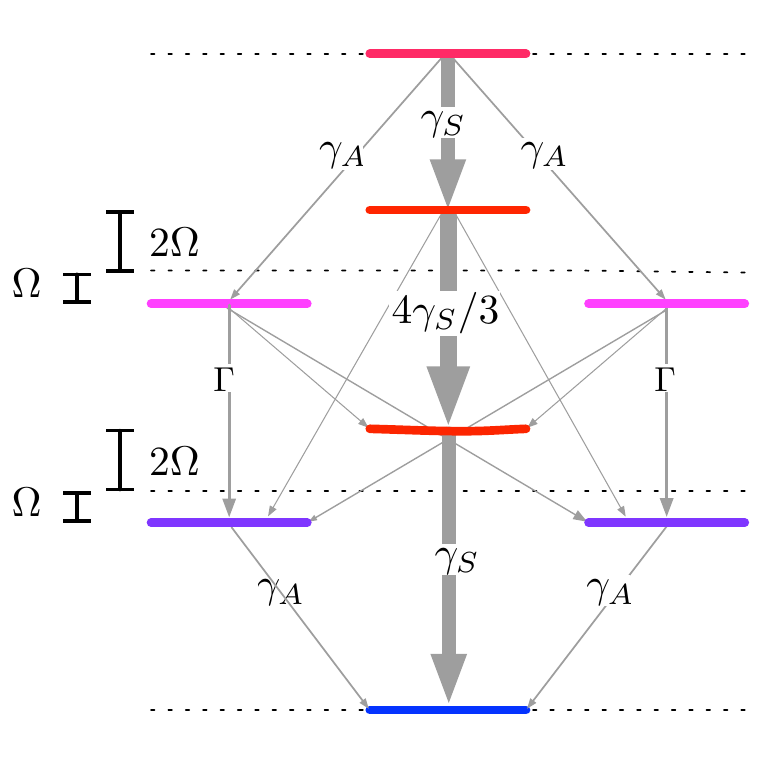}
 \caption{\emph{Energy levels and decay channels for three equidistant atoms}. Results of the diagonalization of both the Hamiltonian and the Liouvillian for three atoms in an equidistant triangle configuration. The dipole-dipole shifts of levels are depicted with the corresponding decay channels and rates. Further details can be found in Ref.~\cite{ostermann2012cascaded}.}
 \label{fig3}
\end{figure}

For three atoms there is still only one choice of $m=1$. However, the resulting state is somewhat more complicated in both coupled and uncoupled bases, i.e.

\begin{displaymath}
\begin{aligned}
\ket{p_3^{(1)}} &= \frac{-1}{2\sqrt{2}}\left(\ket{eee}+\ket{ggg}+\ket{egg}+\ket{gee}\right) \\
& +\frac{1}{4\sqrt{2}}\left(1+i\sqrt{3}\right)\left(\ket{eeg}+\ket{geg}\right) \\
& +\frac{1}{4\sqrt{2}}\left(1-i\sqrt{3}\right)\left(\ket{ege}+\ket{gge}\right) \\
\end{aligned}
\end{displaymath}
\begin{displaymath}
\begin{aligned}
\dots &=\frac{1}{2\sqrt{2}}\left(|\frac{3}{2},\frac{3}{2}\rangle+|\frac{3}{2},-\frac{3}{2}\rangle\right) \\
&+ \frac{\sqrt{{3}}}{4} \left(|\frac{1}{2},\frac{1}{2},1\rangle+|\frac{1}{2},-\frac{1}{2},1\rangle\right) \\
&+i\frac{\sqrt{{3}}}{4}\left(|\frac{1}{2},\frac{1}{2},2\rangle-|\frac{1}{2},-\frac{1}{2},2\rangle\right)
\end{aligned}
\end{displaymath}

Above we have used the short form for the tensor products in the uncoupled basis and an additional index in the coupled basis. The complete label of a state (different from the ones with $J=N/2$) in the coupled basis as used here is $\ket{J,M,\alpha}$, where as usual $0\leq J\leq N/2$ and $|M| \leq J$. In the symmetric subspace, characterized by $J=N/2$ (with states on the surface of the Bloch sphere), there are $N+1$ states. The additional index $\alpha$ is needed in order to distinguish among degenerate states inside the Bloch sphere (note that there is a certain unitary freedom in how the change of basis is performed, i.e. in how the collective degenerate states are defined). These other states that lie inside the Bloch sphere (equal in number to $2^N-(N+1)$), we loosely dub asymmetric states. For the three-particle example, as seen in Fig.~\ref{fig3}, states in the middle belong to the symmetric subspace. There are four such states with maximal $J=3/2$, and therefore $2^3-4=4$ asymmetric states inside the sphere. Since there are only two combinations of $J=1/2$ and $M=\pm1/2$, the remaining states are degenerate and therefore distinguished by an additional index $\alpha=1,2$. These asymmetric state are depicted in Fig.~\ref{fig3} on the sides and correspond to $\ket{1/2,\pm1/2,\alpha}$ in the expression for $\ket{p_3^{(1)}}$.

It is however obvious that the number of asymmetric states grows drastically with $N$ and so does the degeneracy. Consequently, the expressions for the protected states become vastly more complicated for larger $N$ making it necessary to tackle the problem numerically.

For atoms in an equidistant triangle configuration where all mutual decay rates and couplings are equal and specified by $\gamma$ and $\Omega$, respectively, one can again simply use the transformation that diagonalizes the Hamiltonian to diagonalize the Liouvillian as well. The resulting states with their corresponding decay rates are depicted in Fig.~\ref{fig3}. Thee phase-spread transformation that leads to the protected state $\ket{p_3^{(1)}}$ simply ensures that the system's evolution mostly runs through the states on the side, characterized by
smaller decay rates $\gamma_A$.

\section{Results for larger systems}

We are now in the position to extend our investigations to larger systems in various configurations. First, we show results for six atoms in a chain, where the separation is varied and the scan over different rotations (i.e. over all possible sets of $\varphi^m_j$) is performed. We then explain the obtained results by taking a close look at the collective decay properties as derived from a diagonalization of the Liouvillian and find scaling laws for the characteristic timescale of the most protected subspace consistent with the numerical results. Then, we show that the performance of PRT can beat the typical $1/\sqrt{N}$ scaling. Finally, we extend our numerics to a cube configuration of eight atoms which should be the building block for understanding the application of this method to dense 3D lattices.

\subsection{1D chain configuration}

\begin{figure}[t]
\includegraphics[width=0.83\columnwidth]{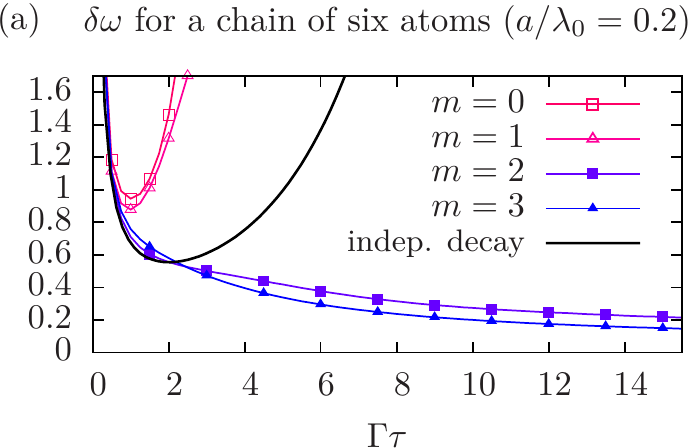}
\\
\includegraphics[width=0.83\columnwidth]{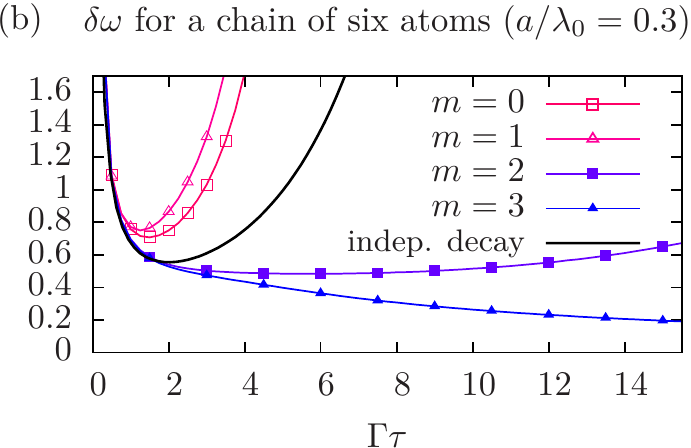}
\\
\includegraphics[width=0.83\columnwidth]{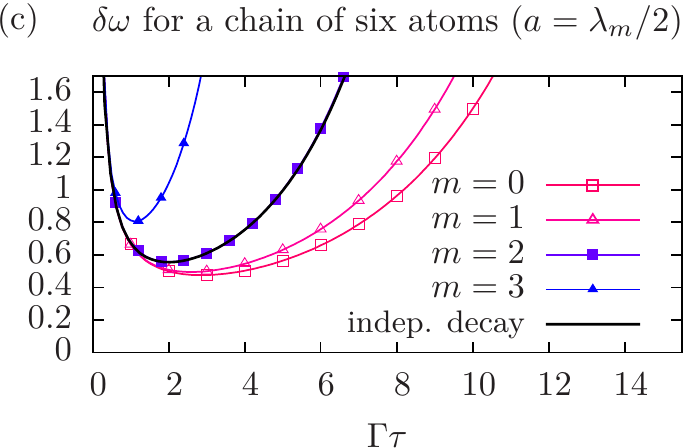}

\caption{\emph{Numerical investigations for a 1D chain of six atoms} Numerical results for the sensitivity as a function of $\tau$ for a 1D chain of six atoms separated by $a=0.2 \lambda_0$ in a), by $a=0.3 \lambda_0$ in b) and by half of the magic wavelength in c). The different curves correspond to independent decay (solid line), SRT with $m=0$ (empty squares) and PRT with $m=1$ (empty triangles), $m=2$ (filled squares) and $m=3$ (filled triangles). In c) the independent decay overlaps with the curve for $m=2$.} \label{fig4}
\end{figure}

To begin with, we consider a linear chain of six atoms separated by various lattice constants $a$ and subject to first SRT and than to PRT. We numerically compute the minimum sensitivity as a function of $\tau$ and scan over all possible rotation indexes $m$. The results are plotted in Fig.~\ref{fig4} for separations of $0.2 \lambda_0$, $0.3
\lambda_0$ and for the magic wavelength. The obtained curves are compared to the independent atom case (shown in black in all plots).

As seen in Fig.~\ref{fig4}a and Fig.~\ref{fig4}b, for distances smaller than $\lambda_0/2$, there is at least one $m$ for which the corresponding PRT method gives results better than SRT. More surprising and promising at the same time, the optimal PRT performs even better than the independent atom case. The immediate conclusion is that one can use such techniques to turn cooperative decay into an advantage instead of treating it as a detrimental effect. For distances larger than $\lambda_0/2$ (as illustrated in Fig.~\ref{fig4}c), the SRT beats any PRT we used for a fairly simple reason. At these distances the symmetric states are subradiant. Therefore, SRT naturally leads the system  to subspaces which are more protected from the environment.

From Fig.~\ref{fig4}a and Fig.~\ref{fig4}b we notice that the $m=3$ rotation performs best. More generally, as seen in the following subsections, the optimal PRT scheme seems to always be the one characterized by a maximum $m=\lfloor N/2 \rfloor$. Such rotations effectively create non-radiative subunits of atom pairs within the chain (exact for even $N$ and an approximation for odd $N$ where one atom is unpaired). This seems to agree with the mechanism described in Ref.~\cite{dorner2013noise} as well.

\begin{figure*}[t]
 \includegraphics[width=1.78\columnwidth]{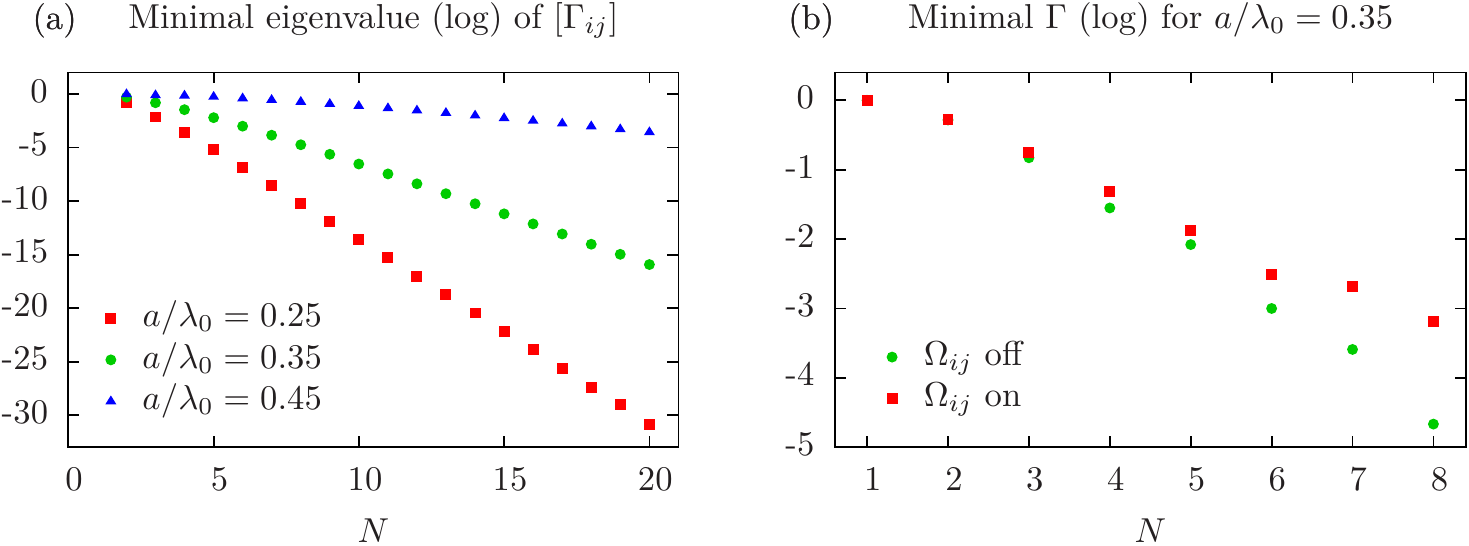}
 \caption{\emph{Subradiant behavior for increasing N}. a) Minimum decay rate (or eigenvalue) obtained from $\left[ \Gamma_{ij} \right]$ with increasing atom number $N$ in a linear chain configuration for different spacings $a$. Smaller distances show a close to exponential and drastic decrease of $\Gamma_{min}$ with increasing $N$. b) Scaling of the $\Gamma_{min}$ obtained via the population accumulation method (as detailed in the text) with and without coherent dipole-dipole energy exchange.}
 \label{fig5}
\end{figure*}

\subsection{Diagonal decay channels - scaling laws}

A key property for the improved performance of dense ensembles under PRT is the occurrence of subradiant states. To get some physical insight into the behavior of these states with distance and particle number, we perform a diagonalization of the decay matrix for $N$ particles in a linear chain configuration. This is done by a unitary matrix $T$, such that
\begin{align}
\Gamma &= T \, D_\Gamma \, T^{-1}, 
\label{evalues}
\end{align}
where $D_\Gamma$ is a diagonal matrix containing the eigenvalues of the decay rate matrix $\left[ \Gamma_{ij} \right]$, which we label $\lambda_i$ for $i=1,...N$. With this, we can write the connection to collective ladder operators as
\begin{align}
\sigma_i^\pm &=: \sum_{k=1}^N T_{ik}\Pi_k^\pm \label{sigma_trafo}.
\end{align}
Using \eqref{sigma_trafo} in \eqref{L_def} we obtain a diagonal form for the Liouvillian that shows a breakdown of the decay process into $N$ different channels, i.e.
\begin{align}
\mathcal{L}[\rho] &=
\sum_{k=1}^N\frac{\lambda_k}{2}(2\Pi_k^-\rho\Pi_k^+ - \Pi_k^+\Pi_k^-\rho - \rho\Pi_k^+\Pi_k^-).
\label{L_final}
\end{align}

After establishing a description of the decay via independent decay channels, let us now investigate the scaling of the corresponding rates with $N$. Results of the numerical diagonalization of the decay matrix $[\Gamma_{ij}]$ are illustrated in Fig.~\ref{fig5}a for $a=0.2 \lambda_0$, $a=0.35 \lambda_0$ and $a=0.45 \lambda_0$. There, the logarithm of the minimum eigenvalue $\Gamma_{min}$ (normalized with respect to $\Gamma$) is plotted against $N$. A closer and closer to exponential scaling emerges as $a/\lambda_0$ becomes smaller.

Having  identified that there are decay channels with exponentially close to zero rates (with increasing $N$), the natural question is: does the system end up in such subspaces characterized by almost perfect protection from the environment? To this end, we simulate population accumulation dynamics, where the system is initialized in the fully inverted state and the population of the ground state is monitored. It is safe to assume that in the long
time limit all but the channel with the very lowest decay rate will have damped out fully. Therefore, the population of the ground state will have the following approximate analytical form for large times,
\begin{equation}
p_{G}(t)\approx 1-e^{-\Gamma_{min}t} \label{GS_approx}.
\end{equation}
The results are plotted in Fig.~\ref{fig5}b as green circles, where $\Omega_{ij}=0$ is assumed. The values obtained perfectly overlap with the predicted values from Fig.~\ref{fig5}a (green circles). However, we have also investigated the effect of coherent dipole-dipole energy exchange  on such dynamics and found the red squares line in Fig.~\ref{fig5}b. In the realistic case where $\Omega_{ij}\neq0$, the Hamiltonian and Liouvillian cannot be  diagonalized simultaneously and the system does not evolve to the fully protected subspace but to a combination of slowly decaying subspaces. The resulting scaling with increasing $N$ is however still quite steep and close to an exponential.

\subsection{Optimal sensitivity via protected method - scaling laws}

\begin{figure}[t]
 \includegraphics[width=0.88\columnwidth]{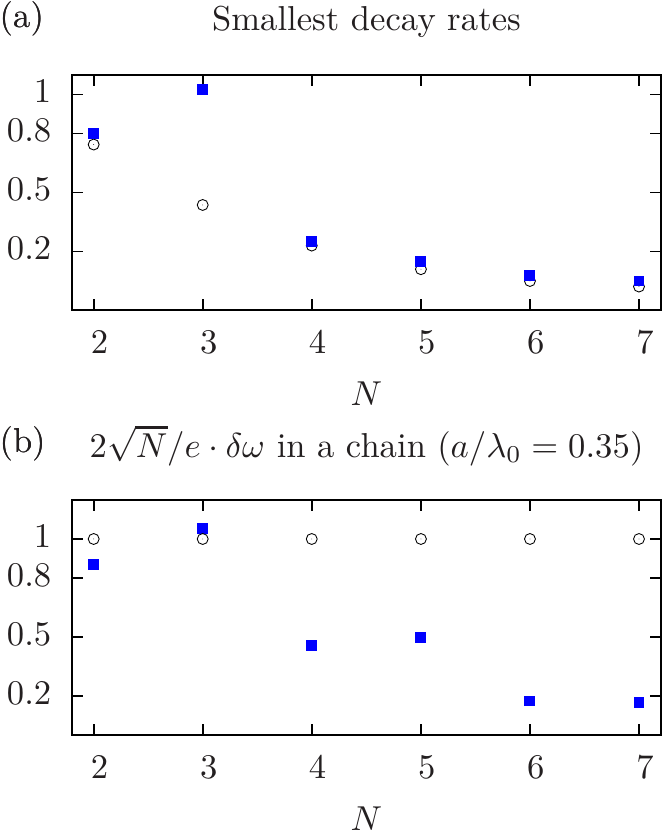}
 \caption{\emph{Scaling laws}. a) Scaling of the extrapolated inverse timescale $2/\tau_{opt}$ with increasing $N$ (squares) compared to the theoretical scaling of $\Gamma_{min}$ (circles). b) Scaling of the minimum sensitivity (times $2 \sqrt{N}/e$) obtained via PRT with particle number (always for the PRT with $m=\lfloor N/2\rfloor$). The circles show the normal scaling of SRT on independent atoms $2 \sqrt{N}/e \cdot \delta \omega =1$. For an even particle number, the rotation with $m=N/2$ corresponds to a configuration where the system is omposed of non-radiative atom pairs. For odd particle number, there is one unpaired spin and the resulting sensitivity is roughly the one obtained for $N-1$ atoms
(except for small systems where the effect of the unpaired atom is substantial).}
 \label{fig6}
\end{figure}

We are now in the position to extract scaling laws for the minimum sensitivity with atom number from numerical investigations of PRT on 1D lattices. First, we extract the optimal interrogation times $\tau_{opt}$ from the sensitivity curves such as those plotted in Fig.~\ref{fig4}. A simple fit of $2/\tau_{opt}$ with the minimum decay rate predicted theoretically for $a=0.35 \lambda_0$ (as read from Fig.~\ref{fig5}a shows a good agreement with increasing $N$ (except for $N=3$ for PRT with $m=1$). The results are shown in Fig.~\ref{fig6}a. The conclusion is that, for long interrogation times, the system subjected to PRT is indeed mainly restricted to a protected subspace governed by the smallest theoretically predicted decay rate.

More importantly, we have analyzed the behavior of the normalized optimal sensitivity $(2\sqrt{N}/e)\delta \omega$ with increasing $N$ and compared it to the typical scaling for independently decaying atoms (shown as a constant function valued $1$ in Fig.~\ref{fig6}b). The immediate conclusion is that PRT does indeed beat the usual scaling using independent ensembles with atoms in coherent spin states and suggests that even the improved scaling introduced by the use of spin squeezed states might be outperformed. However, extended numerical investigations are needed at this point and such an extrapolation will be deferred to a future publication. The chains of even and odd number behave differently, owing to the aforementioned fact that the PRT with $m=\lfloor N/2\rfloor$ is the optimal one for $a=0.35 \lambda_0$. For even $N$ the sensitivity outperforms the standard one as soon as $N>1$ given that every two neighboring atoms are paired into non-radiative cells when PRT is applied. For odd $N$ there is an extra unpaired dipole that seems to strongly influence the results when $N$ is small and will lead to the same $\delta \omega$ (as for the previous even integer) for large $N$.

\subsection{3D cube configuration}

\begin{figure}[t]
\includegraphics[width=0.79\columnwidth]{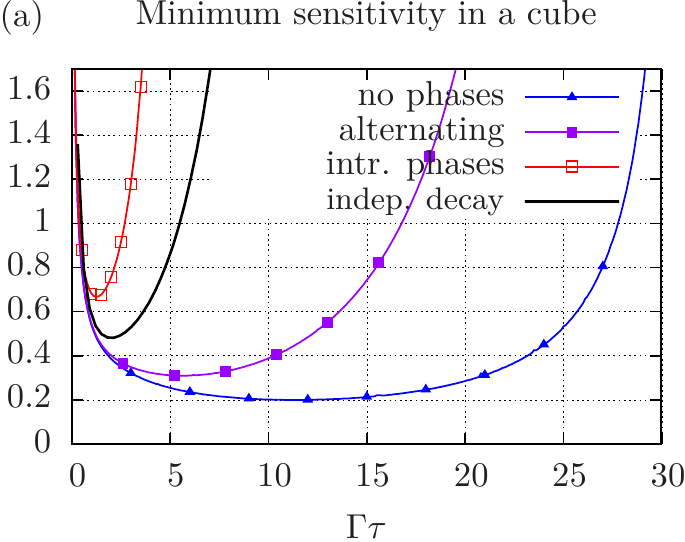}
\\
\includegraphics[width=0.80\columnwidth]{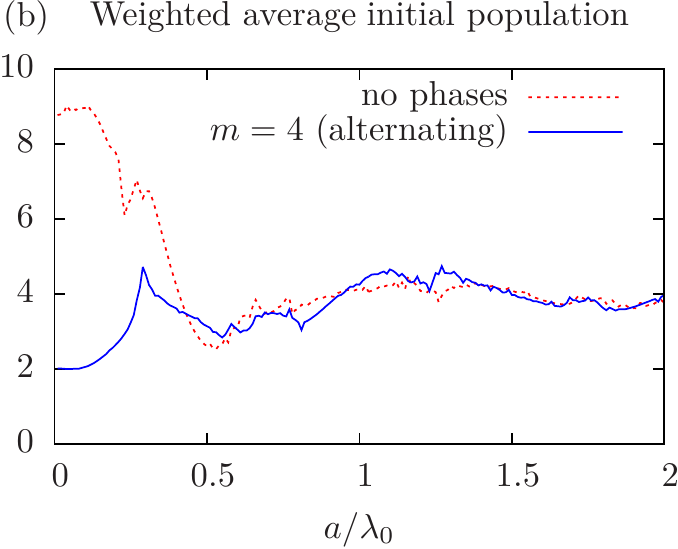}
\caption{\emph{Numerical investigations for the cube configuration}. a) Frequency sensitivity for the cube configuration in a magic wavelength lattice. b) Weighed average of decay rates with corresponding populations for SRT compared to PRT as a function of interatomic distance in a cube. As obvious from the plot, PRT outperforms SRT for distances roughly less
than $\lambda_0/2$.} \label{fig7}
\end{figure}

As a further step towards a generalized view of a 3D configuration this section lays out the properties of a unit cell of a cubic lattice, where eight atoms reside in the corners of a cube. Here, the atoms are trapped equidistantly (lattice constant $a$), while their dipoles point into the direction that is transverse to the propagation direction of the excitation and readout laser
pulses. Thus, different coupling strengths emerge, whereas the dominating contribution is still the nearest-neighbor distance with the dipole moment and the vector connecting the respective pairs drawing an angle of $\theta = \pi/2$. Note that symmetry renders all eight particles equivalent here.

Since, in such a configuration, a direct laser excitation of all eight atoms with the same phase, is not possible, this setup offers a solid testing ground for a quasi-automatic phase imprinting due to finite distances. In a typical situation an excitation pulse would reach one face of the cube, i.e. four atoms, with some phase $\varphi$, while the other four atoms would receive a phase of $\varphi + ka$, where $k$ is the laser's wavenumber and $a$ denotes
the length of the cube, as mentioned above. This, of course, can become arbitrarily complicated, if one allows for the cube to be addressed from any angle, where then each atom could obtain a distinct phase, simply because of the free propagation of the laser pulses between them.

In Fig.~\ref{fig7}a we depict the minimum sensitivity as a function of time for a lattice constant of $a \approx 0.58 \lambda_0$, corresponding to $^{87}Sr$ in a magic wavelength lattice. We observe, that a Ramsey scheme, where every atom receives the same phase outperforms any other phase imprinting by a landslide. This might seems a bit counter-intuitive at first, as one is lead to assume that this situation is the standard Ramsey technique. Yet, a closer look reveals that due to the geometrically induced implicit phase imprinting, the above mentioned second face of the cube needs to pick up an extra phase of $-ka$, so that both faces, i.e. every atom in the cubic sample, indeed possesses the same imprinted phase. At the
magic wavelength distance, the lowest order nearest-neighbor dissipative coupling has a negative value, thus favoring as many pairs of equal phase as possible. As mentioned before, the majority of the couplings is constituted by nearest neighbor pairs, but there are also couplings in the planar and cubic diagonal, which becomes quite evident when looking at the sensitivity for an alternating phase distribution, i.e. every nearest neighbor pair is separated by a phase of exactly $\pi$. Here, a tradeoff between the next-neighbor and diagonal couplings can be observed as the closest couplings decrease the sensitivity due to a negative sign and a phase difference of $\pi$ while the diagonal ones, which are also negative in sign, yet possess no phase difference, increase the sensitivity.

Finally, the SRT sensitivity is obtained by including the implicit phase imprinting caused by a laser pulse that hits one face of the cube first, propagates further and hits the second face with an extra phase of $ka$. At magic wavelength distance this amounts to a phase of approximately $1.16 \pi$, which clearly yields the worst sensitivity as the contributions from the pairs with an unfavorable overall sign outweigh the advantageous ones.

To sum up, for the cubic unit cell, where an implicit geometric phase imprinting has been reversed, it is the blue line (filled triangles) of Fig.~\ref{fig7}a an experimental setup should strive for. This line competes against the SRT line (red, empty squares), which carries the geometrically induced phase difference.

To investigate this elementary building block a little further, Fig.~\ref{fig7}b illustrates the weighted average lifetimes of the initial Ramsey state as a function of the lattice constant for
various phase distributions. The average lifetimes are calculated as
\begin{equation}
\Gamma_{av} = \sum_{j=1}^{2^N} \Gamma_j \left| \left \langle \psi_j \vert \psi_0 \right \rangle \right|^2.
\end{equation}
For very small lattice constants we observe that the PRT yields the lowest average lifetime, while for larger distances having no phase difference between the individual atoms gives better results. Now, this does not necessarily mean that SRT beats PRT at those distances, since, as discussed above, SRT suffers from an implicit imprinting of a phase induced by the sample's geometry.

\section{Conclusions}
Despite the common expectation that pairwise interactions and collective dynamics will introduce shifts and noise to ultrahigh precision spectroscopy setups in dense ensembles, we have shown, that using appropriate intermediate preparation steps, these effects cannot only be minimized but sometimes even used to improve the signal to noise ratio for Ramsey type measurements. Transferring excitation to the so-called protected subspaces prevents errors which cannot be corrected by common rephasing pulse schemes. An important example is the prevention of superradiant decay by a population transfer to subradiant states. Surprisingly, the lifetime of these subradiant states grows very fast with particle density and number, which is reflected in the scaling of both the minimum sensitivity and the maximally allowed interrogation times. While the main focus of this paper is the case of ensembles of cold atoms coupled via dipole-dipole interaction through the electromagnetic vacuum, the idea of using protected subspaces to improve precision spectroscopy can be extended to more general cases of engineered baths. Enhancing the interaction of atoms by coupling to a highly confined field mode ~\cite{zoubi2010hybrid,holland2010} induces long range mutual interactions between any pair of atoms yielding even stronger effects. Recently, analogous implementations using NV-centers or superconducting qubits coupled to CPW transmission lines or resonators showed surprisingly strong effects~\cite{Sandner2010strong, mlynek2014observation, lalumiere2012cooperative, lalumiere2013input}.

In principle, our method in the most general form requires single particle control of the excitation phase. Luckily, in many cases of experimental realizations of such a generalized Ramsey method the required phase pattern has a lot of regularity and symmetries, which can be used to simplify the procedure.  As a first guess one can think of an automatic phase imprinting achieved by the sample's geometry, where the phase front of a plane wave laser hits each element of a regular lattice with a different phase $\exp (i k r_i)$, with $k$ being the wavenumber of the laser and $r_i$ denoting the positions of the atoms. Addressing a linear chain transversally at right angle from the side leads to an equal phase for all particles. By tilting the laser and thus introducing an angle $\alpha$ between the laser's propagation direction and the elongation of the chain the relative excitation phase can be tuned as $\varphi_j = k(j-1)a \cdot \cos(\alpha)$. Alternatively, a magnetic field gradient applied for a prescribed time, resulting in a spatial gradient of the difference in splitting of $\left| g \right \rangle$ and $\left| e \right \rangle$ among the individual two-level emitters, will facilitate the accumulation of a relative phase between the atoms much in the form desired in our scheme. Phase gradients could also be engineered by the differential light shift of off-resonant laser fields. In principle, these phases can be even tailored in 3D.  Finally, an implementation in the framework of engineered baths, e.g. with superconducting qubits coupled to CPW transmission lines, could also be realized. Here one has indeed individual spin control. A more thorough discussion on practical considerations has been provided in Ref.~\cite{ostermann2013protected}.

\section*{Acknowledgments}
We acknowledge the use of the QuTiP open-source
software~\cite{QuTip2013} to generate Fig.~\ref{fig1} and
Fig.~\ref{fig2}. Support has been received from DARPA through the
QUASAR project (L.~O. and H.~R.) and from the Austrian Science Fund
(FWF) via project P24968-N27 (C.~G.).

\end{document}